\documentclass{amsart}
\numberwithin{equation}{section}
\newcommand{\norm}[1]{\left\Vert#1\right\Vert}

\newcommand{\R}{\mathbb R}
\newcommand{\C}{\mathbb C}

\newcommand{\HH}{\mathfrak{H}}
\newcommand{\D}{\mathcal{D}}
\newcommand{\N}{\mathcal{N}}
\newcommand{\la}{\lambda}
\newcommand{\af}{a^{ }_F}
\newcommand{\afd}{a^\dag_F}
\newcommand{\afy}{a^{ }_{F^{-1}}}
\newcommand{\afdy}{a^\dag_{F^{-1}}}
\newcommand{\fn}{\phi^{ }_n}
\newcommand{\bbeta}{\mbox{\boldmath $\beta$}}
\begin{document}

\title{Representations of Coherent States in Non-orthogonal Bases}
\author{S. Twareque Ali$^\dag$, \ \ R. Roknizadeh$^{\dag\dag}$\ \ and \ \
M. K. Tavassoly$^{\dag\dag}$}
\address{$^\dag$Department of Mathematics and Statistics,  Concordia
University, Montr\'eal, Qu\'e\-bec, Canada H4B IR6}%
\address{$^{\dag\dag}$Department of Physics, University of Isfahan, Isfahan,
Iran}

\begin{abstract}
Starting with the canonical coherent states, we demonstrate that all the so-called
nonlinear coherent states, used in the physical literature, as well as large classes
of other generalized coherent states, can be obtained by  changes of bases in the
underlying Hilbert space. This observation leads to an interesting duality between
pairs of generalized coherent states, bringing into play a Gelfand triple of
(rigged) Hilbert spaces. Moreover, it is shown that in each dual pair of families
of nonlinear coherent states, at least one family is
related to a (generally) non-unitary projective representation of the
Weyl-Heisenberg group, which can then be thought of as characterizing the dual pair.
\end{abstract}
 \maketitle
\parskip = 6pt

\section{Introduction}\label{sec-intro}
We begin with the well-known {\em canonical coherent states\/} (CCS), $\vert z \rangle$. In the physical
literature (see, e.g., \cite{AAG-book,klauskag, perel}), these are written
in terms of the so-called Fock basis $\vert n\rangle , \; n= 0, 1, 2, \ldots , \infty$
(or {\em number states\/}):
\begin{equation}
 \vert z \rangle =\N(|z|^2)^{-1/2}\sum_{n=0}^\infty\frac{z^n}{\sqrt{n!}}\;\vert n \rangle,\qquad
 \forall z\in\C\; ,
 \label{CCS}
\end{equation}
where the normalization constant, $\N(|z|^2 )= e^{z^2}$, is chosen so as to ensure that
$\langle z \vert z \rangle =1$. The basis vectors  $\vert n\rangle$ are orthonormal in the
underlying Hilbert space, often termed a {\em Fock space\/}. However, in this paper we shall
use a somewhat more general notation and write,
\begin{equation}\label{CCS-gen}
  \vert z \rangle = \eta_z = \N(|z|^2)^{-1/2}\sum_{n=0}^\infty\frac{z^n}{\sqrt{n!}}\;\fn,\quad
 \forall z\in\C\; ,
\end{equation}
defined as vectors in an abstract (complex, separable) Hilbert space $\HH$, for which the vectors
$\fn$ form an orthonormal basis:
\begin{equation}\label{onb-gen}
  \langle\phi_n|\phi_m\rangle_{\HH}=\delta_{nm},\qquad
  n,m=0,1,2,\cdots,\infty\; .
\end{equation}
The so-called {\em non-linear} coherent states are then defined (see, e.g., \cite{manmarsuza})
by replacing the $n!$ in the denominator following
the summation sign  in (\ref{CCS-gen}) by $x_n! := x_1 x_2 x_3 \ldots x_n$,
where $x_1, x_2 , x_3 , \ldots , $ is a sequence of non-zero
positive numbers and, by
convention, $x_0 ! = 1$. Thus, one obtains, the vectors
\begin{equation}\label{non-lin-CS}
 \eta^{\text{nl}}_z = \N_{\text{nl}}(|z|^2)^{-1/2}\sum_{n=0}^\infty\frac{z^n}{\sqrt{x_n!}}\;\fn\; ,
\end{equation}
where again $\N_{\text{nl}}(|z|^2)$ is an appropriate normalizing constant.
Of course, these are only defined for $z \in \D$, where $\D$ is the open domain in the complex plane
defined by $\vert z \vert < L$, with $L^2 = \lim_{n\rightarrow \infty}x_n$ (provided, of course,
that this limit exists and is non-zero). It is our intention to prove in this
paper that such a family of non-linear coherent states can be obtained via a linear transformation on the
Hilbert space $\HH$, which will amount to replacing the orthonormal set $\{\fn\}_{n=0}^\infty$ by, in general,
a non-orthogonal basis. Under appropriate restrictions, the inverse transformation leads to a {\em dual}
family of non-linear coherent states. This duality is related to a Gelfand triple
\cite{gelvil} of (rigged) Hilbert spaces.
Furthermore, just as the canonical coherent states (\ref{CCS-gen}) can also be defined as the orbit
of a single vector under a projective, unitary representation of the Weyl-Heisenberg group:
\begin{equation}
\eta_z = D(z)\phi_0, \qquad D(z) = e^{\overline{z}a - za^\dag}\; ,
\label{unit-proj-rep}
\end{equation}
it will emerge that in a dual pair of families non-linear coherent states, at least one
family is the orbit of a projective, {\em non-unitary} representation of this same group.
It will also be demonstrated, in particular, that  the well-known
{\em photon-added} states \cite{agartar,roymeh} and the {\em binomial states}
\cite{fufengsol} can also be obtained by such a linear
transformation on $\HH$. However, in these two cases, the non-linear coherent states constructed using
the resulting non-orthogonal bases,  again turn out to be canonical
coherent states and indeed, it is possible to characterize a fairly general class of transformations under
which such a situation prevails.

   It ought to be mentioned at this point that the fact that non-linear coherent states are related to a
choice of a new scalar product on the Hilbert space, has been observed before \cite{becadema,manmarsuza}.
Similarly, the existence of a generalized displacement like operator, related to non-linear coherent states
has been studied earlier \cite{royroy}. However, we unify all these concepts by a systematic
application of a certain class of linear transformations on the underlying Hilbert space. The
resultant appearance of a duality among families of nonlinear coherent states
and of a Gelfand triple in this context, as well as the connection with non-unitary
representations of the Weyl-Heisenberg group, has apparently not been noticed before.

\section{The general setting}\label{sec-genset}

  The primary object for this discussion will be an abstract Hilbert space $\HH$.
Let $T$ be an operator on this space with the properties
\begin{enumerate}

\item $T$ is densely defined and closed; we denote its domain by $\D(T)$.

\item $T^{-1}$ exists and is densely defined, with domain $\D(T^{-1})$.

\item The vectors $\phi_n\in\D(T)\cap\D(T^{-1})$ for all $n$ and there exist non-empty
      open sets $\D_T$ and $\D_{T^{-1}}$ in $\mathbb C$ such that
      $\eta_z \in \D(T), \forall z \in \D_T$ and $\eta_z \in \D(T^{-1}), \forall z \in \D_{T^{-1}}$.
\end{enumerate}
Note that condition (1) implies that the operator $T^\ast T=F$ is self adjoint.

Let
\begin{equation}\label{dual-transf}
\phi_n^F    :=   T^{-1}\phi_n\; ,  \qquad
\phi_n^{F^{-1}} :=  T\phi_n\; , \qquad n =0, 1, 2, \ldots , \infty\; ;
\end{equation}
we define the two new Hilbert spaces:
\begin{enumerate}

  \item $\HH_F$, which is the completion of the set $\D(T)$ in the scaler
product
\begin{equation}\label{F-space}
  \langle f|g\rangle_F  =  \langle f|T^\ast T g\rangle_\HH
     = \langle f|Fg\rangle_\HH.
\end{equation}
The set $\{\phi_n^F\}$ is orthonormal in $\HH_F$ and the map $\phi
\longmapsto T^{-1}\phi ,\; \phi \in \D(T^{-1})$ extends to a {\em unitary}
map between $\HH$ and $\HH_{F}$.
If both $T$ and $T^{-1}$ are bounded, $\HH_F$ coincides with $\HH$ as a set. If $T^{-1}$ is
bounded, but $T$ is unbounded, so that  the spectrum  of
$F$ is bounded away from zero, then $\HH_F$ coincides with
$\D(T)$ as a set.

\medskip

  \item $\HH_{F^{-1}}$, which is the completion of $\D(T^{\ast -1})$ in the
  scaler product
\begin{equation}
  \langle f|g\rangle_{F^{-1}}  =   \langle f|T^{-1}T^{\ast-1}
  g\rangle_\HH
  = \langle f|F^{-1}g\rangle_\HH.
\label{F-inv-space}
\end{equation}
  The set $\{\phi_n^{F^{-1}}\}$ is orthonormal in $\HH_{F^{-1}}$
and the map $\phi\longmapsto T\phi ,\; \phi \in \D(T)$ extends to a {\em unitary}
map between $\HH$ and $\HH_{F^{-1}}$. If $T > I$,
the spectrum of $F$ is bounded away from zero; then $F^{-1}$ is
bounded and one has the inclusions
\begin{equation}\label{gelf-triple}
  \HH_F\subset\HH\subset\HH_{F^{-1}}\; .
\end{equation}
\end{enumerate}

We shall refer to the spaces $\HH_{F}$ and $\HH_{F^{-1}}$ as a {\em dual pair} and when
(\ref{gelf-triple}) is satisfied, the three
spaces $\HH_{F}$, $\HH$ and $\HH_{F^{-1}}$ will be called a {\em Gelfand triple\/} \cite{gelvil}.
(Actually, this is a rather simple example of a Gelfand triple, consisting only of a triplet
of Hilbert spaces \cite{antintr}).

Let $B$ be a (densely defined) operator on $\HH$ and $B^\dag$ its
adjoint on this Hilbert space. Assume that $\D(B)\subset\D(F)$. Then unless $[B,F]=0$,
the adjoint of $B$, considered as an operator on  $\HH_F$ and  which we denote by $B^\ast_F$, is
different from $B^\dag$. Indeed,
\begin{eqnarray*}
\langle f|Bg\rangle_F &=& \langle f|FBg\rangle_\HH
  = \langle B^\dag F f|g\rangle_\HH
  = \langle F F^{-1}B^\dag F f|g\rangle_\HH\\
&=& \langle F^{-1}B^\dag F f|Fg\rangle_\HH
  = \langle F^{-1}B^\dag F f|g\rangle_F,\qquad \forall
f,g\in\D(F)\; .
\end{eqnarray*}
Thus
$$
  B^\ast_F=F^{-1}B^\dag F\; .
$$
On $\HH$ we take the operators $a,a^\dag,N=a^\dag a$:
\begin{equation}\label{osc-alg1}
a\phi_n=\sqrt{n}\phi_{n-1},\quad
a^\dag\phi_n=\sqrt{n+1}\phi_{n+1},\quad N\phi_n=n\phi_n \; .
\end{equation}
These operators satisfy:
\begin{equation}\label{osc-alg2}
  [a,a^\dag]=1,\quad [a,N]=a,\quad [a^\dag,N]=-a^\dag\; .
\end{equation}
On $\HH_F$ we have the transformed operators:
\begin{equation}\label{osc-alg3}
  \af=T^{-1}aT,\quad a^\dag_F=T^{-1}a^\dag T,\quad N_F=T^{-1}NT\; .
\end{equation}
These operators satisfy the same
 commutation relations as $a, a^\dag $ and  $N$ :
\begin{equation}\label{osc-alg4}
  [\af,a^\dag_F]=1,\quad [\af,N_F]=\af,\quad [a^\dag_F, N_F]=-a^\dag_F\; .
\end{equation}
Also on $\HH_F$
\begin{equation}\label{osc-alg5}
\af\phi_n^F=\sqrt{n}\phi_{n-1},\quad
a^\dag_F\phi_n^F=\sqrt{n+1}\phi_{n+1}^F,\quad
N_F\phi_n^F=n\phi_n^F\; .
\end{equation}
 Clearly, considered as operators on
$\HH_F$, $\af$ and $a^\dag_F$ are adjoints of each other and indeed they are just the
unitary transforms on $\HH_F$ of the operators $a$ and $a^\dag$ on $\HH$. On the
other hand, if we take the operator $\af$, let it act on $\HH$
and look for its adjoint on $\HH$ under this action, we obtain by (\ref{osc-alg3}) the
operator $a^\sharp=T^\ast a^\dag T^{\ast-1}$ which, in general, is different
from $a^\dag_F$ and also $[\af,a^\sharp]\neq I$, in general. In an analogous manner, we shall define the
corresponding operators $\afy ,  \afdy$, etc., on $\HH_{F^{-1}}$.

   We thus obtain three unitarily equivalent sets of operators: $a, a^\dag ,  N$, defined on $\HH$,
$\af, \afd ,  N_F$, defined on $\HH_F$ and $\afy, \afdy ,  N_{F^{-1}}$ defined on $\HH_{F^{-1}}$.
On their respective Hilbert spaces, they define under commutation the standard oscillator Lie algebra.
On the other hand, if
they are all considered as operators on $\HH$, the algebra generated by them and their adjoints
on $\HH$ (under commutation) is,
in general, very different from the oscillator algebra and could even be an infinite dimensional Lie
algebra.

Writing $A=\af$, $A^\dag=a^\sharp$, both considered as operators
on $\HH$, if they satisfy the relation
\begin{equation}\label{def-osc-alg}
  AA^\dag-\lambda A^\dag A=C(N)
\end{equation}
where $\lambda\in\R_\ast^+$ is a constant and $C(N)$ is a
function of the operator $N$, then the three operators $A$,
$A^\dag$, $H=(1/2)(AA^\dag+A^\dag A)$ are said to generate a {\it
generalized oscillator algebra} or {\it deformed oscillator
algebra} \cite{borzov95}. Note that on $\HH$, $A$ and $A^\dag$ are adjoints of
each other.

\section{Construction of coherent states}\label{sec-CS-cons}
Consider the vectors
\begin{equation}\label{F-CS}
  \eta_z^F=T^{-1}\eta_z=\N(|z|^2)^{-1/2}\sum_{n=0}^\infty\frac{z^n}{\sqrt{n!}}\;\phi_n^F
\end{equation}
on $\HH_F$. These are the images of the $\eta_z$ in $\HH_F$ and
are the normalized canonical coherent states on this Hilbert space (recall that the
vectors $\phi_n^F$ are orthonormal in $\HH_F$). Similarly,
define the vectors
\begin{equation}\label{F-inv-CS}
  \eta_z^{F^{-1}}=T\eta_z=\N(|z|^2)^{-1/2}\sum_{n=0}^\infty \frac
    {z^n}{\sqrt{n!}}\;\phi_n^{F^{-1}}\; ,
\end{equation}
as the CCS $\eta_z$ unitarily transported from $\HH$ to $\HH_{F^{-1}}$.

  We would now like to consider the  $\eta_z^F$ as being vectors
in $\HH$ and similarly the vectors $\eta_z^{F^{-1}}$ also as
vectors in $\HH$. To what extent can we then call them
(genralized) coherent states? Specifically, we would like to find an orthonormal basis
$\{\psi_n\}_{n=0}^\infty$ in $\HH$ and a transformation $w=f(z)$
of the complex plane to itself such that:
\begin{itemize}
  \item[$(a)$] we could write,
\begin{equation}\label{transf-nlCS}
  \eta_z^F=\zeta_w=\N^\prime (|w|^2)^{-1/2}\; \Omega (w)\sum_{n=0}^\infty\frac{w^n}{\sqrt{[x_n!]}}\;\psi_n\; ,
\end{equation}
where $\N^\prime$ is a new normalization constant, $\Omega (w)$ is a phase factor and
$\{x_n\}_{n=1}^\infty$ is a sequence of non-zero positive numbers, to be
determined;

\medskip

\item[$(b)$] there should exist a measure $d\lambda(\rho)$ on $\R^+$, such
that with respect to the measure
$d\mu(w,\overline{w})=d\lambda(\rho)\; d\vartheta$ (where $w=\rho
e^{i\vartheta}$) the resolution of the identity,
\begin{equation}\label{resolid}
  \int_\D
  |\zeta_w\rangle\langle\zeta_w|\N^\prime(|w|^2)d\mu(w,\overline{w})=I\; ,
\end{equation}
would hold on $\HH$ (as is the case with the canonical coherent states).
Here again, $\D$ is the domain of the complex plane,
$\D=\{w\in \C \;\vert\; \vert w\vert < L\}$,
where $L^2 =\lim_{n\to\infty}x_n$.
\end{itemize}
A general answer to the above question may be hard to find. But
we present below several classes of examples, all physically motivated, for which the above
construction can be carried out. These include in particular all
the so called non-linear, deformed and squeezed coherent states, which
appear so abundantly in the quantum optical and physical literature (see,
for example, \cite{manmarsuza,odz98,simsudmu1}).

 Whenever the two sets of vectors $\{\eta_z^F\}$ and $\{\eta_z^{F^{-1}}\}$
 form coherent state families in the above sense, we shall call
 them a {\it dual pair}.

 \section{Examples of the general construction}\label{sec-ex}

 \subsection{Example 1}{\em Photon-added and binomial states as bases}

 Let $T$ be an operator such that $T^{-1}$ has the form
\begin{equation}\label{ex1T}
  T^{-1}=e^{\lambda a^\dag}G(a),
\end{equation}
where $\lambda\in\R$ and $G(a)$ is a function of the operator $a$
such that $T$ and $T^{-1}$ satisfy the postulated conditions
(1)-(4) of Section \ref{sec-genset}. (The operator $G(a)$ could, for example, be defined by
taking an entire analytic function $G(z)$ with real coefficients
and non-zero in the finite plane, and then setting
$G(a)\eta_z=G(z)\eta_z$ for all $z\in\C$). It is easily verified
that
\begin{equation}\label{ex1-adj}
e^{\lambda a^\dag}a = (a-\lambda I)e^{\lambda a^\dag}\; , \qquad
e^{\lambda a}a^\dag = (a^\dag+\lambda I)e^{\lambda a}\; ,
\end{equation}
so that
\begin{equation}\label{ex1-adj2}
  e^{\lambda a^\dag}G(a)=G(a-\lambda I)e^{\lambda a^\dag}.
\end{equation}
From this we compute the two transformed operators $\af$ and
$a^\dag_F$ on $\HH_F$ ($F=T^\ast T=e^{-\la a}G(a^\dag)^{-1}G(a)^{-1}e^{-\la
a^\dag}$) to be:
\begin{equation}\label{ex1-adj3}
\af  =  T^{-1}aT=a-\la I\; , \qquad
\afd =  T^{-1}a^\dag T=G(a-\la I)a^\dag G(a-\la I)^{-1}\; .
\end{equation}
Thus, since $a$ commutes with $G(a-\la I)$, we obtain
\begin{equation*}
  [a^{ }_F,a^\dag_F] = G(a-\la I)[a,a^\dag]G(a-\la
  I)^{-1}= 1\; ,
\end{equation*}
as expected. The two operators $A=\af$ and $A^\dag=T^\ast a^\dag
T^{\ast^{-1}}$, defined on $\HH$, are
\begin{equation}\label{ex1-adj4}
  A=a-\la I,\quad A^\dag=a^\dag-\la I
\end{equation}
which of course are adjoints of each other. Moreover, in this case
\begin{equation}\label{ex1-adj5}
  [A,A^\dag]=I\; ,
\end{equation}
so that the oscillator algebra remains unchanged.

Since, by (\ref{ex1-adj}),
\begin{equation*}
  ae^{-\la a^\dag}=e^{-\la a^\dag}(a-\la I),
\end{equation*}
we see that
\begin{equation}\label{ex1-adj6}
T=G(a)^{-1}e^{-\la a^\dag}=e^{-\la a^\dag}G(a-\la I)^{-1}\; .
\end{equation}
  Thus we get the corresponding operators,
  \begin{equation}\label{ex1-adj7}
  \afy = TaT^{-1}=a+\la I\; , \qquad
  \afdy  =  Ta^\dag T^{-1}=G(a)^{-1}a^\dag G(a)\; ,
  \end{equation}
on $\HH_{F^{-1}}$. Once again we obtain $[\afy,\afdy]=1$ and
similarly for the operator $A'=\afy=a+\la I$ and its adjoint
$A'^\dag=a^\dag+\la I$ on $\HH$.

We now define the vectors
\begin{equation}\label{phot-add-bin}
  \phi^F_n=T^{-1}\phi_n=e^{\la a^\dag}G(a)\phi_n\; ,
\end{equation}
which form an orthonormal set in $\HH_F$, and build the
corresponding canonical coherent states
\begin{equation}\label{phot-add-CS}
  \eta^F_z=\N(|z|^2)^{-1/2}\sum_{n=0}^\infty\frac{z^n}{\sqrt{n!}}\phi_n^F=e^{\la
  a^\dag}G(a)\eta_z\; ,
\end{equation}
on $\HH_F$. Considering these as vectors in $\HH$, and taking account of the
fact that
\begin{equation*}
a e^{\lambda a^\dag}G(a)= e^{\lambda a^\dag}G(a)(a+\lambda I),
\end{equation*}
we see that
\begin{equation}\label{phot-add-bin2}
a \eta^{F}_{z}= (z+\la)\eta^{F}_{z}.
\end{equation}
Thus, up to a constant factor, $\eta^F_z$ is just the canonical coherent state on
$\HH$ corresponding to the point $(z+ \lambda)\in \C$ (note that
since the canonical coherent states  can be obtained as solutions to a first order
differential equation, $(x+d/{dx}) \eta_{z}=z\eta_{z}$, the
solution is unique, up to a constant, for each $z\in \C$, i.e., to each $z \in \mathbb C$, there
corresponds exactly one vector $\eta$ such that $a\eta = z\eta$). We write, therefore,
$$
 \eta_z^{F}= C(\lambda , z )\sum_{n=0}^\infty\frac{(z+\lambda)^n}{\sqrt{n!}}\;\fn\; ,
$$
where the constant $C(\lambda , z )$ can be
computed by going back to (\ref{phot-add-CS}). Indeed, we have,
\begin{eqnarray*}
\eta_z^F & = & e^{\lambda a^\dag} G (a)\eta_z = G(z)e^{\lambda a^\dag}\eta_z\\
     & = & G(z)e^{- \frac {\vert z \vert^2}2}e^{\lambda a^\dag}e^{za^\dag}\phi_0
             = G(z)e^{- \frac {\vert z \vert^2}2}e^{\frac {\vert z + \lambda\vert^2}2}\eta_{z + \lambda}\\
     & = & G(z)e^{\lambda (\Re (z) + \frac {\lambda}2 )}\eta_{z + \lambda}\; .
\end{eqnarray*}
Thus, we get $C(\lambda , z ) = G(z)e^{- \frac {\vert z \vert^2}2}$ and
\begin{equation}\label{phot-add-bin3}
 \eta_z^F = G(z)e^{- \frac {\vert z \vert^2}2}\;\sum_{n=0}^\infty\frac{(z+\lambda)^n}{\sqrt{n!}}\;\fn
        = G(z)e^{\lambda (\Re (z) + \frac {\lambda}2 )}\eta_{z + \lambda}\; .
\end{equation}

Comparing (\ref{phot-add-bin3}) with (\ref{transf-nlCS}) and writing
$\eta^{F}_{z}=\zeta^{}_{z+\la}$, we find that $w=z+\la$,
$x^{}_{n}=n$ and $\psi^{}_{n}=\fn$. Furthermore, $\N^\prime (\vert w \vert^2 ) = e^{\vert z \vert^2}
\vert G(z)\vert^{-2}$ and $\Omega (w ) = e^{i\Theta (w )}$, where we have written
$G(z ) = \vert G(z)\vert e^{i\Theta (w )}$.   It is remarkable that in this example
while $\eta^{F}_{z}$ is written in (\ref{phot-add-CS})
in terms of a non-orthonormal basis $\{\phi_n^F\}_{n=0}^\infty$,
when these vectors are considered as constituting a basis for $\HH$, its transcription in terms
of the orthonormal basis $\{\phi_n\}_{n=0}^\infty$
only involves a shift in the variable $z$ and no change in the
components.

It is now straightforward to write down a resolution of identity,
following the pattern of the canonical coherent states. Indeed,
writing $w=z+\la=\rho e^{i\theta}$, we have (on $\HH$),
\begin{equation}\label{resolid2}
\int\!\!\int_{\C}|\zeta_{w}\rangle \langle
\zeta_{w}|\N_{F}(|w|^{2})d\mu(w,\overline{w})=I \; , \qquad
d\mu(w,\overline{w})=\frac{e^{-\rho^{2}}}{\pi}\rho d\rho d\theta\; .
\end{equation}

The dual CS $\eta^{F^{-1}}_{z}$ are obtained by replacing the
$\phi^{F}_{n}$ in (\ref{phot-add-bin}) by
$\phi^{F^{-1}}_{n}=T\fn=G(a)^{-1}e^{-\la a^\dag} \fn$. But since
$G(a)^{-1}e^{-\la a^\dag}=e^{-\la a^\dag }G(a- \la I)^{-1}$, we
have
\begin {equation}\label{phot-add-bin-dual1}
\phi^{F^{-1}}_{n}=e^{-\la a^ \dag}G(a-\la I)^{-1}\fn.
\end {equation}
Hence, using the same argument as with the $\phi^F_{n}$, we
arrive at
\begin {equation}\label{phot-add-bindual2}
\eta^{F^{-1}}_{z}= G(z - \lambda)^{-1}e^{- \frac {\vert z \vert^2}2}\;\sum_{n=0}^{\infty}
\frac{(z-\la)^{n}}{\sqrt{n!}}\;\fn   =
G(z - \lambda )^{-1}e^{-\lambda (\Re (z) - \frac {\lambda}2 )}\eta_{z - \lambda}\; .
\end {equation}
Thus, in the present case (up to normalization),
the dual pair of states $\eta^F_z$ and $\eta^{F^{-1}}_z$ is obtained simply by replacing
$\la$ by $-\la$.

 It is clear now that the above construction can be carried out
 for any operator $T^{-1}$ which satisfies the commutation
 relation
 \begin {equation}\label{phot-add-comm}
[a, T^{-1}]=\la T^{-1},\qquad \la\in\mathbb R
 \end {equation}
with $a$.

Two particular cases of the operator  $T^{-1}$ in
(\ref{ex1T}) are of special interest. In the first instance take $G(a)=I$, so that
$T^{-1}=e^{\la a^\dag}$. The vectors $\phi^{F}_{n}=T^{-1}\fn$ may
easily be calculated. Indeed we get
\begin {equation}\label{phot-add1}
\phi^{F}_{n}=\sum_{k=0}^{\infty}\frac{(\la a^\dag)^{k}}
{k!}\fn=e^{\frac{\la^{2}}{2}}\frac{{a^\dag}^{n}}{\sqrt{n!}}\eta^{}_{\la}\; ,
 \end {equation}
which (up to  normalization) are the well-known {\em photon-added coherent states\/}
of quantum optics \cite{agartar, roymeh}.
Hence in this case we write $\phi^{F}_{n}=\phi^{\text{pa}}_{\la, n}$. We
denote the corresponding coherent states by $\eta^{\text{pa}}_{\la,
z}$ and note that
\begin{eqnarray}\label{phot-add2}
\eta^{F}_{z}:=\eta^{\text{pa}}_{\la,z} &=&
\N(|z|^{2})^{-\frac{1}{2}}\sum_{n=0}^{\infty}\frac{z^{n}}{\sqrt{n!}}\phi^{\text{pa}}_{\la ,n}
= \N(|z|^{2})^{-\frac{1}{2}}e^{\frac{\la^{2}}{{2}}}e^{za^\dag}\eta^{}_{\la}\nonumber\\
&=& e^{\la(x+\frac{\la}{2})}\eta^{}_{z+\la}\; ,
\end{eqnarray}
where $x=\Re(z)$. Clearly if $\la\longrightarrow 0$, then
$\eta^{\text{pa}}_{\la,z}\longrightarrow\eta^{}_{z}$. The dual set of
coherent states, $\eta^{F^{-1}}_{z}$ are obtained by replacing $\la$ by $-\la$ so that
the states $\eta^{\text{pa}}_{\la, z}$ and $\eta^{\text{pa}}_{-\la,
z},\; z \in \mathbb C$,  are in duality, and we have the interesting relation,
\begin {equation}\label{phot-add3}
 \langle\eta^{\text{pa}}_{-\la,
z}|\eta^{\text{pa}}_{\la, z} \rangle_{\HH}=e^{-\la(\la+2iy)}.
\end {equation}

On $\HH^{}_{F}$ we have the creation and annihilation
operators (see (\ref{ex1-adj3})),
\begin {equation}\label{phot-add4}
a^{}_{F}=a-\la I,\quad a^\dag_{F}=a^\dag\; ,
\end {equation}
which are adjoints of each other on $\HH^{}_{F}$, but clearly not so
on $\HH$. However, on $\HH$ we have the two operators $A$ and
$A^\dag$ as in (\ref{ex1-adj4}):
$$ A=a-\la I,\quad A^\dag=a^\dag-\la I\; . $$

   As the second particular case of (\ref{ex1T}), we
take $\la=0$ and $G(a)=e^{\mu a},\;  \mu\in\Re,$ i.e., $T^{-1}=e^{\mu
a}$. The basis vectors are now
\begin {equation}\label{bin1}
\phi^{F}_{n}=e^{\mu
a}\fn=\sqrt{n!}\sum_{k=0}^{n}\frac{\mu^{n-k}}{\sqrt{k!}(n-k)!}\fn=\frac{(a^\dag+\mu
I)^{n}}{\sqrt{n!}}\;\phi_{0}.
\end{equation}
These states have also been studied in the quantum optical literature \cite{fufengsol} and
in view of the last expression in (\ref{bin1}), we shall call
them {\em binomial states} and write $\phi^F_n=\phi^{\text{bin}}_{\mu , n}$. The
coherent states, built out of these vectors as basis states, are:
\begin{eqnarray}\label{bin2}
\eta^F_z:=\eta^{\text{bin}}_{\mu,z} &=& e^{\mu
a}\eta_z=e^{-|z|^2/2}\sum_{n=0}^\infty\frac{z^n}{\sqrt{n!}}\; \phi^{\text{bin}}_{\mu, n}\nonumber\\
&=& e^{\mu x-|z|^2/2}\sum_{n=0}^\infty\frac{z^n}{\sqrt{n!}}\;\fn.
\end{eqnarray}
The dual CS are simply $\eta^{\text{bin}}_{-\mu,z}$ and
\begin{equation}\label{bin3}
  \langle\eta^{\text{bin}}_{-\mu,z}|\eta^{\text{bin}}_{\mu,z} \rangle=1.
\end{equation}
The creation and annihilation operators on $\HH_F$ are:
\begin{equation}\label{bin4}
  \af=a\; ,\qquad \afd=a^\dag+\la I\; ,
\end{equation}
while the other two operators on $\HH$ are:
\begin{equation}\label{bin5}
  A=a\; ,\qquad A^\dag=a^\dag\; .
\end{equation}
The operators (\ref{bin4})
have been studied, in the context of non-self-adjoint Hamiltonians in
\cite{becadema,bedeszaf}.
Again, it is remarkable that the coherent states $\eta^{\text{bin}}_{\mu,z}$ are exactly the
canonical coherent states, $\eta^{ }_z$, up to a factor.

   Before leaving this example, a further point ought to be made in
connection with the two basis sets
$\{\phi_{\la, n}^{\text{pa}}\}^\infty_{n=0}$ and
$\{\phi_{\mu,n}^{\text{bin}}\}^\infty_{n=0}$, consisting of the photon-added
coherent states and the binomial states, respectively. The set
$\{\phi_{\la,n}^{\text{pa}}\}^\infty_{n=0}$ is orthonormal with respect to the
operator
\begin{equation}\label{phot-bin-op1}
F_{\text{pa}} =e^{-\la a}e^{-\la a^{\dag}}=e^{\la^2/2}e^{-\sqrt{2}\;\la Q}\; ,
\end{equation}
where
\begin{equation}
Q=\frac{1}{\sqrt{2}}(a+a^\dag)
\label{posop}
\end{equation}
is the usual position operator. Thus, we have
\begin{equation}\label{phot-bin-op2}
  e^{\la^2/2}\langle\phi^{\text{pa}}_{\la,n}|e^{-\sqrt{2}\la
  Q}\phi^{\text{pa}}_{\la,m}\rangle_\HH=\delta_{mn}\; .
\end{equation}
 The operator $e^{-\sqrt{2}\; \la Q}$ has a completely continuous
spectrum ranging from $0$ to $\infty$. On the other hand, the set
$\{\phi^{\text{bin}}_{\mu,n}\}^\infty_{n=0}$ is  orthonormal with respect
to the operator
\begin{equation}\label{phot-bin-op3}
  F_{\text{bin}}=e^{-\mu a^\dag}e^{-\mu a}=e^{-\mu^2/2}e^{-\sqrt{2}\;\mu Q}\; ,
\end{equation}
so that
\begin{equation}\label{phot-bin4}
e^{-\mu^2/2}\langle\phi^{\text{bin}}_{\mu,n}|e^{-\sqrt{2}\mu
Q}\phi^{\text{bin}}_{\mu,m}\rangle_\HH=\delta_{mn}\; .
\end{equation}
Since for $\la=\mu$, $e^{\la^2}F_{\text{bin}}=F_{\text{pa}}$, i.e., the two
operators only differ by a constant, the vectors $\phi^\text{pa}_{\la,n}$
and $\phi^{\text{bin}}_{\la,n}, \; n=0,1,2,\cdots,$ must be unitarily related, up
to a constant. Indeed, since in this case,
\begin{equation*}
  \phi^{\text{pa}}_{\la,n}=e^{\la a^\dag}\phi_n,\qquad {\textrm{and}}\qquad
  \phi^{\text{bin}}_{\la,n}=e^{\la a}\phi_n,
\end{equation*}
we easily get
\begin{equation}\label{phot-bin5}
  \phi^{\text{pa}}_{\la,n}=e^{\la^2/2}\;V \phi^{\text{bin}}_{\la,n},\qquad n=0,1,2,\cdots,
\end{equation}
where $V$ is the unitary operator
\begin{equation}\label{phot-bin6}
  V=e^{-i\sqrt{2}\la P}\; ,\qquad P=\frac{a-a^\dag}{i\sqrt{2}}\; .
\end{equation}

\subsection{Example 2}{\em Re-scaled basis states and nonlinear CS}

For the next general class of examples, let the operator $T^{-1}$ have the form
\begin{equation}\label{ex2T}
  T^{-1}:=T(N)^{-1}=\sum_{n=0}^\infty \frac{1}{t(n)}|\fn\rangle\langle\fn|
\end{equation}
where the $t(n)$ are real numbers, having the properties:
\begin{enumerate}
  \item $t(0)=1$ and $t(n)=t(n')$ if and only if $n=n'$\; ;
  \item $0<t(n)<\infty$\; ;
  \item the finiteness condition for the limit
\begin{equation}\label{Tfin}
 \lim_{n\to\infty}\left[\frac{t(n)}{t(n+1)}\right]^2\cdot\frac{1}{n+1}=\rho<\infty
\end{equation}
holds.
\end{enumerate}
This last condition implies that the series
\begin{equation}\label{Tfin2}
  \sum_{n=0}^\infty\frac{r^{2n}}{[t(n)]^2n!}:=S(r^2)
\end{equation}
converges for all $r<L=1/\sqrt{\rho}$. The operators $T$ and
$F$ are now
\begin{equation}\label{ex2TF}
 T := T(N)  =  \sum_{n=0}^\infty t(n)|\fn\rangle\langle\fn|\; , \qquad
 F := F(N) = \sum_{n=0}^\infty t(n)^2|\fn\rangle\langle\fn|\; .
\end{equation}
Let us define a new operator $f(N)$,  by its action on the basis vectors.
\begin{equation}\label{nonlinf}
  f(N)\fn :=\frac{t(n)}{t(n-1)}\fn =f(n)\fn,
\end{equation}
then
\begin{equation}\label{nonlinf2}
  t(n)=f(n)f(n-1)\cdots f(1):=f(n)!.
\end{equation}
Thus we have the transformed, non-orthogonal basis vectors
\begin{equation}\label{nonlinf3}
  \phi^F_n=\frac{1}{t(n)}\fn=\frac{1}{f(n)!}\fn \; ,
\end{equation}
so that if $\psi=\sum_{n=0}^\infty c_n\fn$ and
$\psi'=\sum_{n=0}^\infty c'_n\phi'_n$ are vectors in $\HH$ which
lie in the domain of $T^{-1}$, then their scalar product in
$\HH_F$ is
\[
\langle\psi|\psi'\rangle_F=\sum_{n=0}^\infty\frac{\overline{c}_nc'_n}{[f(n)!]^2}.\]
We shall call the vectors (\ref{nonlinf3}) {\em re-scaled basis states\/}.

      The coherent states $\eta^F_z$ are now:
\begin{equation}\label{nonlinCS}
  \eta^F_z=\N(|z|^2)^{-1/2}\sum_{n=0}^\infty\frac{z^n}{\sqrt{n!}}\phi^F_n,
\end{equation}
which, as vectors in $\HH_F$ are well defined and normalized for
all $z\in \C$. However, when considered as vectors in $\HH$ and
rewritten as:
\begin{equation}\label{nonlinCS2}
  \eta^F_z=\N(|z|^2)^{-1/2}\sum_{n=0}^\infty\frac{z^n\fn}{f(n)!\sqrt{n!}},
\end{equation}
are no longer normalized and defined only on the domain (see
\ref{Tfin} and \ref{Tfin2}),
\begin{equation}\label{nonlindom}
  \D=\left\{z\in\C\Big| |z|<L=\frac{1}{\rho}\right\}.
\end{equation}
The operators $\af$ and $\afd$ act on the vectors $\phi^F_n$
 as
\begin{equation}\label{nonlinops}
  a_F\phi^F_n=\sqrt{n}\phi^F_{n-1}\; ,\qquad
  \afd\phi^F_n = \sqrt{n+1}\phi_{n+1}^F\; .
\end{equation}
The operator $A=a_F$, considered as an operator on $\HH$ and its
adjoint $A^\dag$ on $\HH$ act on the original basis vectors $\fn$ in the manner,
\begin{equation}\label{nonlinops2}
  A\fn =  f(n)\sqrt{n}\phi_{n-1}\; , \qquad
  A^\dag\fn  =  f(n+1)\sqrt{n+1}\phi_{n+1}\; ,
\end{equation}
and thus, we may write, in an obvious notation,
\begin{equation}\label{nonlinops3}
  A=af(N)\; ,\qquad A^\dag=f(N)a^\dag\; ,
\end{equation}
as operators on $\HH$.

Thus, up to normalization, the CS defined in (\ref{nonlinCS2}) are
the well-known {\it non-linear coherent states} of quantum optics
\cite{manmarsuza}.

    As a specific physical example of such a family of coherent states,
we might mention the function
$f(n)=L^{(0)}_n(\eta^2)[(n+1)L^{(0)}_n(\eta^2)]^{-1}$, where
$L^m_n(x)$ are generalized Laguerre polynomials and $\eta$ is the
so-called Lamb-Dicke parameter. These states appear as the stationary
states of the centre of mass motion of a trapped and bichromatically
laser driven ion, far from the Lamb-Dicke regime \cite{filvog}.

  The dual coherent states $\eta^{F^{-1}}_z$ which, as vectors in the Hilbert space $\HH_{F^{-1}}$,
will be well-defined vectors in $\HH$ only if
\begin{equation}\label{nonlindualdom}
  \lim_{n\to\infty}\left[ \frac{t(n+1)}{t(n)}\right]^2\cdot\frac{1}{n+1}=\widetilde{\rho}<\infty.
\end{equation}
In this case we have
\begin{equation}\label{nonlinCSdual}
  \eta^{F^{-1}}_z=\N(|z|^2)^{-1/2}\sum_{n=0}^\infty\frac{f(n)!z^n}{\sqrt{n!}}\;\fn
\end{equation}
and are defined (as vectors in $\HH$) on the domain
\begin{equation}\label{nonlindualdom2}
\widetilde{\D}=\left\{z\in\C\Big|
|z|<\widetilde{L}=\frac{1}{\sqrt{\widetilde{\rho}}}\right\}.
\end{equation}
Equations (\ref{nonlinCSdual}) and (\ref{nonlindualdom2}) should be compared to
(\ref{nonlinCS2}) and (\ref{nonlindom}). We also have
\begin{equation}\label{nonlinduality}
  \langle\eta_z^{F^{-1}}\Big |\eta_z^F\rangle_\HH=1,
\end{equation}
for all $z\in \D\cap\widetilde{\D}$.

A resolution of the identity  of $\HH$ can be obtained in terms of the vectors
$\eta_z^F$ (or $\eta_z^{F^{-1}}$) by solving a moment problem.
Thus, for example, for the vectors (\ref{nonlinCS2}) to satisfy,
\begin{equation}\label{nonlinresolid}
  \int\!\!\int_\D|\eta_z^F\rangle\langle\eta_z^F|\N(|z|^2)d\mu(z,\bar{z})=I\; ,
\end{equation}
where $d\mu(z,\bar{z})=d\la(r)d\theta$, ($z=re^{i\theta}$), the
measure $d\la$ must satisfy the moment conditions
\begin{equation}\label{nonlinmomcond}
  \int_0^L r^{2n}d\la(r)=\frac{\left[f(n)! \right]^2
  n!}{2\pi},\quad n=0,1,2,\cdots \; .
\end{equation}

  As is well known, the most nonclassical
features of nonlinear coherent states lie in their  squeezing,
antibunching and sub-Poissonian properties, which
all depend crucially  on the choice of the nonlinearity
function. These properties have been studied for nonlinear coherent
of the dual type (\ref{nonlinCSdual}) in  \cite{royroy}.

  A highly instructive example of the duality between families of non-linear coherent
states is provided by the Gilmore-Perelomov \cite{gilmore} and Barut-Girardello
\cite{bar-gir71} coherent states, defined for the
discrete series representations of the group $SU(1,1)$. The
Gilmore-Perelomov coherent states can be defined on $\HH$ as:
\begin{equation}\label{gilperCS}
  \eta^{\text{GP}}_z=\N_{\text{GP}}(|z|^2)^{-1/2}\sum_{n=0}^\infty\sqrt{\frac{(2\kappa +n-1)!}{n!}}z^n\;\fn
\end{equation}
where $\N_{\text{GP}}$ is a normalization factor, chosen so that
$\norm{\eta_z^{\text{GP}}}^2_\HH=1$, and the parameter
$\kappa =1,3/2,2, 5/2, \cdots$,
labels the $SU(1,1)$ representation being used.
These coherent states are defined on the open unit disc, $|z|<1$. The
Barut-Girardello coherent states, on the other hand, can be defined (again on
$\HH$) as the vectors
\begin{equation}\label{bargirCS}
  \eta_z^{\text{BG}}=\N_{\text{BG}}(|z|^2)^{-1/2}\sum_{n=0}^\infty\frac{z^n}{\sqrt{n!(2\kappa +n-1)!}}\;\phi_n\; ,
  \quad z \in \C\; ,
\end{equation}
where once more, $\N_{\text{BG}}$ is chosen so that
$\norm{\eta_z^{\text{BG}}}^2=1$. It is now immediately clear that the
operator
\begin{equation}\label{bargir-gilper-trans}
  T(N)=\sum_{n=0}\frac{1}{\sqrt{(2\kappa +n-1)!}}|\fn\rangle\langle\;\fn|\;
\end{equation}
acts in the manner,
\begin{equation}
  \eta_z^{\text{BG}} = \lambda_1 \; T(N)\eta_z\; \qquad \text{and} \qquad
  \eta_z^{\text{GP}} = \lambda_2\;T(N)^{-1}\eta_z\; ,
\label{bargir-gilper-trans2}
\end{equation}
where $\lambda_1$ and $\lambda_2$ are constants,
thus demonstrating the relation of duality between the two sets of coherent states.

   A large class of dual pairs of the above type can be constructed
by starting with the hypergeometric function,
\begin{equation}
 _pF_q (\alpha_1 , \alpha_2 , \ldots , \alpha_p ; \; \beta_1 , \beta_2 , \ldots , \beta_q ; \; x )
       = \sum_{n=0}^\infty \frac {(\alpha_1)_n (\alpha_2)_n \ldots (\alpha_p)_n}
           {(\beta_1)_n (\beta_2)_n \ldots (\beta_q)_n}\; \frac {x^n}{n!}\; ,
\label{hypergeomfcn}
\end{equation}
where the $\alpha_i$ and $\beta_i$ are positive real numbers,  $q$ is an arbitrary positive integer and $p$
is restricted by $q-1 \leq p \leq q +1$. (Here $(\gamma)_n$ is
the usual Pocchammer symbol, $(\gamma)_n = \gamma (\gamma+1)(\gamma+2)\ldots (\gamma+n-1) =
\Gamma (\gamma + n)/\Gamma (\gamma)$). This series converges for all $x \in \R$ if $p = q$ and for all
$\vert x \vert < 1$ if $p = q+1$. Then, going back to the canonical coherent states on $\HH$, we apply to them the
operators
\begin{eqnarray}
 T  :=  T(N) & = & \sum_{n=0}^\infty \left[\frac {(\alpha_1)_n (\alpha_2)_n \ldots (\alpha_p)_n}
           {(\beta_1)_n (\beta_2)_n \ldots (\beta_q)_n}\right]^{\frac 12}\;\vert\fn\rangle\langle\fn\vert\; ,
           \nonumber\\
 T^{-1}  :=  T(N)^{-1} & = & \sum_{n=0}^\infty \left[\frac {(\alpha_1)_n (\alpha_2)_n \ldots (\alpha_p)_n}
           {(\beta_1)_n (\beta_2)_n \ldots (\beta_q)_n}\right]^{-\frac 12}\;\vert\fn\rangle\langle\fn\vert\; .
 \end{eqnarray}
It is then immediate that the corresponding families of coherent states $\{\eta_z^F\}$ and
$\{\eta_z^{F^{-1}}\}$ will be in duality. (Actually, it may be necessary to impose additional restrictions on
the $\alpha_i$ and $\beta_i$, in order to ensure that the coherent states $\{\eta_z^F\}$ and
$\{\eta_z^{F^{-1}}\}$, when defined on $\HH$, satisfy a resolution of the identity \cite{appschill}).

   To conclude this example, we note that from the manner in which the operators $T$ and $T^{-1}$ are defined,
for the re-scaled basis states (see (\ref{ex2T}) and  (\ref{ex2TF})),  we can always
arrange to be in one of the following two situations:
\begin{enumerate}
\item both $T$ and $T^{-1}$ are bounded;
\item $T$ is unbounded but $T^{-1}$ is bounded.
\end{enumerate}
In both cases, (\ref{gelf-triple}) holds, so that we always have a Gelfand triple.

\subsection{Example 3.}{\em Squeezed bases}

Our next  example involves the use of {\em squeezed states} and squeezed bases
(see, for example (\cite{AAG-book,simsudmu1}). Consider the symplectic group,
$\text{Sp}(2, \R )$, consisting of $2\times 2$ real  matrices
$M$ satisfying
\begin{equation}
  M\bbeta M^T = \bbeta\; , \qquad \bbeta = \begin{pmatrix} 0 & 1 \\ -1 & 0 \end{pmatrix}\; .
\label{sympgrp}
\end{equation}
(Note that these matrices can also be characterized by the simple condition, $\text{det}M = 1$,
i.e., $\text{Sp}(2, \R )$ is identical with the group $\text{SL}(2, \R )$, of $2\times 2$ real
matrices of determinant one). An element $M \in \text{Sp}(2, \R )$ has the well-known decomposition
\cite{sugiura},
\begin{equation}
   M = \begin{pmatrix} 1& 0 \\ -v & 1\end{pmatrix}
   \begin{pmatrix} u^{-\frac 12}& 0 \\ 0 & u^{\frac 12}\end{pmatrix}
   \begin{pmatrix} \cos\theta & -\sin\theta \\ -\sin\theta & \cos\theta\end{pmatrix}\; ,
\label{sp2decomp}
\end{equation}
with $v \in \R\;, \;\; u > 0\; , \;\; 0< \theta \leq 2\pi$.
We shall also write,
\begin{equation}
  M(u,v) = \begin{pmatrix} 1& 0 \\ -v & 1\end{pmatrix}
   \begin{pmatrix} u^{-\frac 12}& 0 \\ 0 & u^{\frac 12}\end{pmatrix} =
   \begin{pmatrix} u^{-\frac 12} & 0 \\ -vu^{-\frac 12}  & u^{\frac 12} \end{pmatrix}\; .
\label{sp2decomp2}
\end{equation}

 Next, writing $z = \displaystyle{\frac 1{\sqrt{2}}}(q - i p)$, we introduce the
vector $\mathbf{x}$ and the vector operator $\mathfrak X$:
\begin{equation}
\mathbf{x} = \begin{pmatrix} q \\ p \end{pmatrix}\; , \qquad
\mathfrak X = \begin{pmatrix} Q \\ P \end{pmatrix}\; ,
\label{vecop}
\end{equation}
where $Q$ and $P$ are the position and momentum operators defined in (\ref{posop})
and (\ref{phot-bin6}), respectively. In terms of these quantities
the canonical coherent states (\ref{CCS-gen}) can be rewritten as,
\begin{equation}
\eta_{\mathbf{x}} := \eta_z = U(\mathbf x )\phi_0\;,\qquad  \text{where,}\qquad
  U(\mathbf x ) = \exp [-i\mathbf{x}^T \bbeta\mathfrak X ]\; ,
\label{altCCS}
\end{equation}
and $U(\mathbf x )$ is unitary on $\HH$. If $\begin{pmatrix} Q' \\ P'\end{pmatrix} = M\mathfrak X$,
$M \in \text{Sp}(2, \R )$, then since $[Q', P'] = [Q,P] = iI$, there exists a unitary operator
$U(M)$ on $\HH$ such that (with a slight abuse of notation),
\begin{equation}
  U(M)\mathfrak X U(M)^\dag = M^{-1}\mathfrak X \qquad \text{and} \qquad U(M)U(\mathbf x )U(M)^\dag
     = U(M\mathbf x )\; .
\label{sp2action}
\end{equation}

    Taking $\HH = L^2 (\R , dx )$ and $\phi_0 = \pi^{-\frac 14}e^{-\frac {x^2}2}$, the states,
\begin{equation}
 \eta_{\mathbf x}^{u,v} = U(\mathbf x )U(M(u,v))\phi_0\; , \quad
  (\eta_{\mathbf x}^{u,v})(x) = \left[\frac u\pi\right]^{\frac 14}e^{i(x-\frac q2 ) p}
        e^{-\frac 12 (x -q)(u + iv ) (x -q)}\; ,
\label{sqeez1}
\end{equation}
are generalized Gaussians and for $v=0, u = \frac 1{s^2}$ these are {\em squeezed states\/}.

   For fixed $M(u,v)  \in \text{Sp}(2, \R )$, let $T^{-1} = U(M(u,v))$ and set $\phi^{u,v}_n =
\phi_n^F = U(M(u,v))\phi_n$. We call the resulting basis a {\em squeezed basis\/.} Then
\begin{equation}
\eta^F_z = e^{-\frac {\vert z\vert^2}2}\sum_{n=0}^\infty \frac {z^n}{\sqrt{n!}}\phi_n^{u,v}\; ,
\label{sqCS}
\end{equation}
and since by (\ref{sp2action}),
\begin{eqnarray*}
  U(M(u,v))U(\mathbf x ) &  = &  U(M(u,v) )U(\mathbf x)U(M(u,v))^\dag U(M(u,v))\\
      & = & U(M\mathbf x )U(M(u,v)),
\end{eqnarray*}
we obtain,
\begin{equation}
 \eta_z^F = \eta_{\mathbf x'}^{u,v}\; ,  \quad \text{where} \quad \mathbf x' = M(u,v)\mathbf x\; .
\label{sqCS2}
\end{equation}
Thus, squeezing the basis results in squeezing the coherent states. The dual family of coherent states
consists of the vectors $\eta_{\mathbf x''}^{1/u,-v}$, with $\mathbf x'' = M(1/u,-v)\mathbf x$.
Since $U(M(u,v))$ is unitary on $\HH$, the algebra generated by the operators $A$ and $A^\dag$ is the
same as that generated by $a$ and $a^\dag$.

\section{Some operator algebras}\label{sec-opalg}
  In this Section we take a  closer look at the two sets of operators $a_F , a^\dag_F$ and
$a_{F^{-1}}, a_{F^{-1}}^\dag$ and the algebras generated by them (under commutation), in the special
case when the operators $T$ and $F$  have the forms given in (\ref{ex2TF}). Note that both $T$ and $F$
are positive operators. As noted earlier, on the Hilbert space $\HH_F$ the operators $a_F , a^\dag_F$
are adjoints of each other and satisfy the commutation relation $[a_F , a^\dag_F] = I$, while on the
Hilbert space $\HH_{F^{-1}}$ the operators $a_{F^{-1}}, a_{F^{-1}}^\dag$ are mutual adjoints, satisfying
$[a_{F^{-1}}, a_{F^{-1}}^\dag] = I$. As before, let us write $A = a_F$, when this operator acts on $\HH$ and
similarly we write $A' = a_{F^{-1}}$ to denote the action of $a_{F^{-1}}$ on $\HH$. Since $a_F = T^{-1}aT$
and $a_{F^{-1}}^\dag = Ta^\dag T^{-1}$ and since $T$ and $T^{-1}$ are positive operators, we have for the
adjoint of $A$ on $\HH$,
\begin{equation}
  A^\dag = Ta^\dag T^{-1} = a_{F^{-1}}^\dag\; ,
\label{A-adj}
\end{equation}
and similarly, for the adjoint of $A'$ on $\HH$ we have
\begin{equation}
  A^{\prime\dag} = T^{-1}a^\dag T = a_F^\dag\; .
\label{Aprime-adj}
\end{equation}
Moreover (see (\ref{nonlinops3})),
\begin{equation}
  A = a f(N)\;, \quad A^\dag = f(N)a^\dag\;  \quad \text{and} \quad A' = af(N)^{-1}\; , \quad A^{\prime\dag}
      = f(N)^{-1}a^\dag\; ,
\label{A-aprime-ops}
\end{equation}
with
\begin{equation}
  [A, A^{\prime\dag}] = [A' , A^\dag ] = I\; .
\label{A-aAprime-comm}
\end{equation}
In addition, we have the four other easily verifiable commutation relations,
\begin{eqnarray}
  [A, A^\dag ] & = & f(N+1)^2 (N+1) - F(N)^2 N \;, \nonumber\\
  \left[A', A^{\prime\dag} \right] & = & f(N+1)^{-2} (N+1) - F(N)^{-2} N \; ,\nonumber \\
  \left[A, A'\right] & = & a^2[f(N-1)f(N)^{-1} - f(N-1)^{-1}f(N)\; , \nonumber\\
  \left[A^\dag , A^{\prime\dag}\right] & = &
               [f(N)f(N-1)^{-1} - f(N)^{-1}f(N-1)]a^{\dag 2}\; .
\label{A-aAprime-comm2}
\end{eqnarray}

   Consider now the {\em displacement operators} on $\HH$,
\begin{equation}
 D(z) = e^{za^\dag - \overline{z} a} = U(\mathbf x )\; , \qquad z \in \C\; .
 \label{dispop1}
 \end{equation}
These operators are unitary on $\HH$ and in view of the relation
\begin{equation}
 D(z_1 ) D(z_2 ) = e^{i\Im(\overline{z}_1 z_2)}\; D(z_1 + z_2)\; ,
 \label{WH-projrep}
 \end{equation}
together they realize a unitary projective representation of the Weyl-Heisenberg
group on $\HH$. Moreover,
\begin{equation}
  \eta_z = D(z)\phi_0 = e^{-\frac {\vert z\vert^2}2}\; e^{za^\dag}\phi_0\; .
\label{dispop2}
\end{equation}
The unitary images of $D(z)$ on $\HH_F$ and $\HH_{F^{-1}}$ are,
\begin{equation}
 D_F(z) = T^{-1}D(z)T = e^{za^\dag_F - \overline{z} a_F} \quad \text{and} \quad
 D_{F^{-1}}(z) = TD(z)T^{-1} = e^{za^\dag_{F^{-1}} - \overline{z} a_{F^{-1}}} \; ,
\label{WH-projrep2}
\end{equation}
respectively, again defined for all $z \in \C$ and realizing unitary projective representations
of the Weyl-Heisenberg group on $\HH_F$ and $\HH_{F^{-1}}$, respectively.  Also, just as in
(\ref{dispop2}), we have,
\begin{equation}
\eta_z^F = D_F(z)\phi_0 = e^{-\frac {\vert z\vert^2}2}\; e^{z a^\dag_F}\phi_0\;,  \qquad
\eta_z^{F^{-1}} = D_{F^{-1}}(z)\phi_0 = e^{-\frac {\vert z\vert^2}2}\; e^{za_{F^{-1}}^\dag}\phi_0\; .
\label{dispop3}
\end{equation}
Letting them act  on $\HH$, we write $V(z)$ and $V'(z)$ for these two operators,
so that using (\ref{A-adj}) and (\ref{Aprime-adj}), we have,
\begin{equation}
V(z) := D_F (z) = e^{zA^{\prime\dag} - \overline{z} A}  \quad \text{and} \quad
V'(z) :=  D_{F^{-1}}(z) =  e^{zA^\dag - \overline{z} A'}\; ,
\end{equation}
operators which have been studied in \cite{royroy}.
Thus, as operators on $\HH$,
\begin{equation}
V'(z) = V(-z)^\dag =[V(z)^{-1}]^\dag\;.
\label{contragrep}
\end{equation}
However, on $\HH$ the operator $V(z)$ is only defined for $z \in \mathcal D$, where $\mathcal D$ is the
domain (\ref{nonlindom}), while $V'(z)$ is defined for $z \in \widetilde{\mathcal D}$
(see (\ref{nonlindualdom2})), so that (\ref{contragrep}) only holds on
$\mathcal D \cap \widetilde{\mathcal D}$.  Also, if $z_1 , z_2, z_1 + z_2 \in \mathcal D$ then
we have a relation similar to (\ref{WH-projrep}) for $V(z)$:
\begin{equation}
V(z_1 ) V(z_2 ) = e^{i\Im(\overline{z}_1 z_2)}\; V(z_1 + z_2)\; .
\label{contragrep2}
\end{equation}
Similarly,  if $z_1 , z_2, z_1 + z_2 \in \widetilde{\mathcal D}$ then
we have for $V'(z)$ the analogous relation:
\begin{equation}
V'(z_1 ) V'(z_2 ) = e^{i\Im(\overline{z}_1 z_2)}\; V'(z_1 + z_2)\; .
\label{contragrep3}
\end{equation}
Thus, if $\mathcal D = \C$ (respectively, $\widetilde{\mathcal D} = \C$) then the operators $V(z)$
(respectively, $V'(z)$) define a {\em non-unitary} projective representation
of the Weyl-Heisenberg group on $\HH$. In case $\mathcal D = \widetilde{\mathcal D} = \C$, then
both $V(z)$ and $V'(z)$ realize non-unitary representations of the Weyl Heisenberg group on $\HH$ and
(\ref{contragrep}) implies that these representations are {\em contragredient} to
each other. This could happen, if for example, both $T$ and $T^{-1}$ are bounded operators. Another
possibility could be when $T$ and $T^{-1}$ have the forms:
\begin{eqnarray}
 T & = &\sum_{n=0}^\infty \frac {(\alpha_1)_n (\alpha_2)_n \ldots (\alpha_p)_n}
 {(\beta_1)_n (\beta_2)_n \ldots (\beta_p)_n}\; \vert\fn\rangle\langle\fn\vert\;, \nonumber\\
 T^{-1} & = & \sum_{n=0}^\infty \left[ \frac {(\alpha_1)_n (\alpha_2)_n \ldots (\alpha_p)_n}
 {(\beta_1)_n (\beta_2)_n \ldots (\beta_p)_n} \right]^{-1} \; \vert\fn\rangle\langle\fn\vert\; ,
 \label{contragrep4}
\end{eqnarray}
for real numbers $\alpha_j$ and $\beta_j$. (This corresponds to taking $p$ = $q$ in
(\ref{hypergeomfcn})). But in all cases, one member of a dual pair gives rise to a
non-unitary projective representation of the Weyl-Heisenberg group. In other words, each dual
pair of nonlinear coherent states is characterized by such a representation.

   Finally, we note that the general method which emerges for constructing nonlinear coherent states
is to take the two operators $T$, $D(z)$, defined as in  (\ref{ex2TF}) and (\ref{dispop1}),
a {\em fiducial vector} $\phi_0$,  and then setting
\begin{equation}
  \eta_z^{\text{nl}} = T^{-1}D(z)\phi_0\; .
\label{gen-nonlinCS}
\end{equation}
The set of values of $z$ for which these vectors are defined then depends on $T$. The dual family of
non-linear CS is defined by replacing $T$ by $T^{-1}$. The canonical CS form a self-dual family.

\section*{Acknowledgements}
Part of this work was done while one of the authors (STA) was visiting the Department of Physics,
University of Isfahan. He would like to thank the Department and the University for hospitality.
He would also like to thank the Natural Sciences and Engineering Research Council (NSERC) of Canada
for a research grant.


\begin{thebibliography}{99}

\bibitem[Agarwal(1991)]{agartar} Agarwal, G.K. and Tara, K., {\em Non-classical properties of states
                    generated by the excitations on a coherent state\/}, Phys. Rev {\bf A43}, 492-497 (1991).
\medskip
\bibitem[Ali(2000)]{AAG-book} Ali, S.T., Antoine, J-P. and Gazeau, J-P.,
                   {\em Coherent States, Wavelets and their Generalizations\/}, Springer-Verlag, New York, 2000.
\medskip
\bibitem[Antoine(2002)]{antintr} Antoine, J.-P., Inoue, A. and Trapani, C., {\em Partial *-Algebras
                 and Their Operator Realizations\/}, in {\em Mathematics and Its Applications\/}, vol. 553,
                 Kluwer, Dordrecht, NL, 2002.
\medskip
\bibitem[Appl(2003)]{appschill} Appl, T. and Schiller, D.H., {\em Generalized hypergeometric coherent
                   states\/}, preprint quant-ph0308013 v1 1 Aug 2003.
\medskip
\bibitem[Barut(1971)]{bar-gir71} Barut, A.O. and Girardello, L., {\em New ``coherent'' states associated
                    with non-compact groups\/}, Commun. Math. Phys. {\bf 21}, 41-55 (1971).
\medskip
\bibitem[Beckers(2001)]{becadema} Beckers, J., Cari\~{n}ena, J.F., Debergh, N. and Marmo, G.. {\em Non-hermitian
                    oscillator-like Hamiltonians and $\lambda$-coherent states revisited\/}, Mod. Phys. Lett.
                    {\bf A16}, 91-98 (2001).
\medskip
\bibitem[Beckers(1998)]{bedeszaf} Beckers, J., Debergh, N. and Szafraniec, F.H., {\em A proposal of new sets of
                    squeezed states\/}, Phys. Letters {\bf A243}, 256-260 (1998) and {\bf A246},
                    561 (1998).
\medskip
\bibitem[Borzov(1997)]{borzov95} Borzov, V.V., Damaskinsky, E.V. and Yegorov, S.B., {\em Some remarks on
                   the representations of the
                   generalized deformed oscillator algebra\/}, preprint q-alg/9509022 v1; Zap. Nauch. Seminarov.
                   LOMI {\bf 245}, 80-106 (1997) (in Russian).
\medskip
\bibitem[Filho(1996)]{filvog} Filho, R.L. de Matos and Vogel, W., {\em Nonlinear coherent states\/}, Phys. Rev.
                   {\bf 54A}, 4560-4563 (1996).
\medskip
\bibitem[Fu(2000)]{fufengsol} Fu, H., Feng, Y. and Solomon, A.I., {\em States interpolating between number and
                    coherent states and their interaction with atomic systems\/}, J. Phys. {\bf A33}, 2231-2249
                    (2000).
\medskip
\bibitem[Gelfand(1964)]{gelvil} Gelfand, I.M. and Vilenkin, N. Ya., {\em Generalized Functions\/}, Vol. 4,
                   Academic Press, New York, 1964.
\medskip
\bibitem[Gilmore(1974)]{gilmore} Gilmore, R., {\em Lie Groups, Lie Algebras, and Some of their applications},
                    John Wiley \& Sons, New York (1974).
\medskip
\bibitem[Klauder(1985)]{klauskag}Klauder, J.R and Skagerstam, B.S., {\em Coherent States, Applications in Physics and
                   Mathematical Physics}, World Scientific, Singapore, (1985).
\medskip
\bibitem[Manko(1997)]{manmarsuza} V.I. Man'ko, G. Marmo, E.C.G. Sudarshan and F. Zaccaria, {\em $f$-oscillators and
                   non-linear coherent states\/}, Physica Scripta {\bf 55}, 528-541 (1997).
\medskip
\bibitem[Odzijewicz(1998)]{odz98} Odzijewicz, A., {\em Quantum algebras and $q$-special
                   functions related to coherent states maps of the  disc\/},
                   Commun. Math. Phys. {\bf 192}, 183-215 (1998).
\medskip
\bibitem[Perelomov(1986)]{perel}Perelomov, A.M., {\em Generalized coherent states and their applications},
                   Springer-Verlag, Berlin (1986).
\medskip
\bibitem[A.Roy(1995)]{roymeh} Roy, A.K. and Mehta, C.L. {\em Boson inverse operators and associated coherent states\/},
                   Quantum Semiclass. Opt {\bf 7}, 877-888 (1995).
\medskip
\bibitem[B.Roy(2000)]{royroy} Roy, B. and Roy, P. {\em New nonlinear coherent states and some of their nonclassical
                   properties\/}, J.Opt B: Quantum Semiclass. Opt. {\bf 2}, 65-68 (2000).
\medskip
\bibitem[Simon(1988)]{simsudmu1} Simon, R., Sudarshan, E.C.G.  and Mukunda, N.,  {\em Gaussian
           pure states in quantum mechanics and the symplectic group\/},
           Phys. Rev. {\bf  A37},  3028--3038  (1988).
\medskip
\bibitem[Sugiura(1990)]{sugiura} M. Sugiura, {\em Unitary Representations and Harmonic Analysis: An
               Introduction\/},  North-Holland/Kodansha Ltd., Tokyo (1990).






\end{thebibliography}
\end{document}